\documentclass[a4paper,11pt]{article}
\pdfoutput=1
\usepackage{jheppub} 
\usepackage{lineno}
\usepackage{mathtools}
\usepackage{xcolor}

\title{ \boldmath Constraining Inflation via FIMP dark matter using the $\beta$-function with collider implications}

\author[a]{Sarif Khan,}
\author[b]{Sourov Roy,}
\author[c]{and Ananya Tapadar}

\affiliation[a]{
Department of Physics, Chung-Ang University, Seoul 06974, Korea.
}
\affiliation[b]{
School of Physical Sciences, Indian Association for the Cultivation of Science,\\
Raja Subodh Chandra Mallick Rd, Jadavpur, Kolkata-750032, India
}
\affiliation[c]{Institute of Nuclear Physics, IFJ-PAN, Radzikowskiego 152, 31-342 Krak\'{o}w, Poland}

\emailAdd{sarifkhan@cau.ac.kr}
\emailAdd{tpsr@iacs.res.in}
\emailAdd{atapadar@ifj.edu.pl}

\abstract{ The present study connects inflation and freeze-in type dark matter (DM) within the same setup. Although the observables in these two phenomena lie at vastly different energy scales, they have been properly handled using the RG running of couplings.
For studying DM and inflation, the SM has been minimally extended by introducing an abelian dark gauge symmetry and a dark singlet scalar. In studying inflation, the SM Higgs doublet has been considered 
as the inflaton, which has a non-minimal coupling with the Ricci scalar. 
All inflationary observables have been computed at the horizon exit scale and constrained using the Planck data. 
Moreover, inflationary constraints have revealed strong correlations 
among model parameters, significantly reducing the allowed parameter space.
In particular, in the Higgs mixing angle and BSM Higgs mass plane, only those values that ensure the Higgs quartic coupling remains above $0.18$ are allowed. 
The additional gauge boson serves as a suitable DM candidate, produced via the freeze-in mechanism and stabilised by charge conjugation symmetry.
The upper bound on the DM relic density further shrinks the parameter space 
allowed from inflationary constraints, becoming even narrower if we assume that the present vector DM constitutes the total DM density. 
Since DM interactions are feeble, it remains safe from all terrestrial experimental constraints. Additionally, the feeble dark matter coupling requires the dark Higgs–Ricci scalar non-minimal coupling to be negligible to satisfy Higgs inflation conditions. 
Finally, we have explored collider aspects and found that the trilinear and quartic Higgs vertices deviate from their SM values after incorporating inflation and DM constraints. Therefore, once we measure $\kappa_{3,4}$ at the future collider, we can establish the robustness of the Higgs inflation scenario.
}
\makeatletter
\def\@fpheader{}
\def\@preprint{}
\makeatother
\begin{document}
\maketitle
\flushbottom

\section{Introduction}
\label{sec:intro}

In the present work, we have addressed two important phenomena of our Universe: one is inflation, and the other is dark matter (DM). 
Both of them happen at different scales of the Universe, {\it i.e.}, inflation is an early
Universe phenomenon, whereas DM is a late Universe phenomenon\footnote{UV freeze-in DM \cite{Hall:2009bx, Biswas:2025adi} or gravitational production of DM \cite{Lebedev:2022cic} happens at the high scale.}. Therefore, it is
difficult to connect inflation and DM in one setup and even more difficult
to have the influence of one sector into the other. 
In this work, as we will see later in detail, we have tried to study the effect of DM production and bounds on inflation by making the running of the couplings 
from the low scale to the high scale.
Inflation \cite{Guth:1980zm, Linde:1981mu, Albrecht:1982wi} is a well-accepted phenomenon 
in the early Universe, 
which explains many unsolved issues associated with the standard Big Bang theory. 
The main shortcomings of the standard Big Bang cosmology are the flatness problem, which questions the origin of the highly spatially flat Universe in the early era; the entropy problem which is related to the flatness problem; the horizon problem which questions why causally disconnected regions in the early Universe share the same temperature; and the overproduction of magnetic monopoles in grand unified theory in the early Universe. The introduction of inflation in the early Universe after the Big Bang can address all the aforementioned problems very effectively.
During the inflation era, the Universe went through exponential enhancement
$\left( \frac{a(t_f)}{a(t_i)} \right)^{2} = e^{H_{I} (t_{f} - t_{i})}$
where $H_I$ is the Hubble parameter, which is constant during inflation
and $t_{i(f)}$ is the initial (final) time of inflation.
In explaining inflation, we need inflaton to slow roll in the early Universe, and many different kinds of potentials have been proposed to explain inflation. 
The power-law potential for the inflation field is disfavored by the Planck data 
\cite{Planck:2018vyg}. Other allowed potentials for inflation are the 
Starobinsky potential \cite{Starobinsky:1980te}, 
the $\alpha-$attractor model \cite{Kallosh:2013yoa}
and the non-minimal coupling of scalars to the Ricci scalar 
\cite{Cervantes-Cota:1995ehs, Bezrukov:2007ep}.

In Refs. \cite{Cervantes-Cota:1995ehs, Bezrukov:2007ep}, Higgs inflation using 
the non-minimal coupling was first introduced 
and is followed in the present work. Later, its validity was studied using the running of 
the couplings \cite{Bezrukov:2008ej} and the full Standard Model (SM) bath 
production from inflaton decay, 
considering a high reheating temperature \cite{Bezrukov:2008ut}.
The proposed Higgs inflation using the non-minimal coupling was questioned in 
Refs. \cite{Burgess:2009ea, Barbon:2009ya, Lerner:2009na, Burgess:2010zq,
Hertzberg:2010dc} due to concerns about unitarity violation 
beyond the cutoff scale and naturalness. In Ref. \cite{Bezrukov:2010jz}, 
it was shown that the cutoff scale of the effective theory for Higgs inflation is background field-dependent and the cutoff scale always remains above the field value. Therefore, one can safely continue the study of inflation up to the Planck scale since the cutoff scale of the effective theory always remains above the field value even if
the theory is not UV-complete.

Another long-standing unsolved problem is the DM production mechanisms, its nature, and 
detection strategies in different experiments. 
The most popular DM candidate is the weakly interacting massive particle (WIMP)
which is produced via the freeze-out mechanism \cite{Gondolo:1990dk}. 
The continuous advancement 
of direct detection experiments has severely constrained the WIMP DM parameter 
space. Recent data from LUX-ZEPLIN \cite{LZ:2024zvo} has ruled out many scenarios 
where DM and the SM have direct interactions without any mixing. 
Therefore, it is high time to explore alternatives to WIMP-type DM. 
The alternative popular DM candidate is the feebly interacting massive 
particle (FIMP), which is produced via the freeze-in 
mechanism \cite{McDonald:2001vt, Hall:2009bx}. 
In this mechanism, 
DM is assumed to have zero initial abundance in the early Universe and 
is continuously produced from bath particles until it freezes in. FIMP DM has extremely 
feeble couplings and never achieves thermal equilibrium, making DM detection 
prospects very difficult. On a lighter note, feeble interaction could 
be a possible reason for the non-detection of DM so far.

In the present work, we investigate DM production via the freeze-in mechanism while 
constraining the inflation sector. In this respect, we have extended the SM by an additional
abelian gauge group and a singlet scalar. The gauge boson of $U(1)_D$ is the
DM in our work.
A similar study for WIMP-type DM in the context 
of inflation has been explored recently in Ref. \cite{Khan:2023uii}.
In inflation, we have considered the non-minimal coupling of the SM Higgs with gravity. 
This has an advantage compared to the simplest inflation model,
$m^{2} \phi^2 + \lambda \phi^4$, where one requires $m \sim 10^{13}$ GeV and 
$\lambda \sim 10^{-13}$ for successful inflation \cite{Linde:1983gd}. 
In contrast, for Higgs inflation, where we have $\xi \phi^{2} R$ ($R$ is Ricci scalar and $\xi$ is non-minimal coupling), 
the CMB fluctuation is fixed by the combined factor 
$\lambda/\xi^2$. Therefore, in the case of 
Higgs inflation, we have $\lambda \sim 0.1$, and $\xi$ can be adjusted accordingly to obtain the correct value obtained from CMB.
In our study, we have successfully explained inflation using the two-loop running 
effect of the model parameters from the top-quark pole mass ($m_t$) to 
the Planck scale ($M_{\text{pl}}$). We have varied the model parameters within certain ranges 
and obtained the correct values for the spectral index ($n_s$), tensor to scalar
ratio ($r$), amplitude of the curvature power spectrum ($A_s$) and DM relic density
which are consistent with the recent Planck data \cite{Planck:2018vyg}. These quantities have been measured at the horizon exit or pivot scale, so sometimes they will be referred to with the subscript ``piv".
In particular, we have obtained a sharply anti-correlated allowed region in the 
$M_{h_2}$ and $\sin\theta$ plane, which provides Higgs quartic coupling $(\lambda_H)$ value of 
 in the 
range $0.18-0.25$ at the $m_t$ scale required to keep the SM Higgs quartic 
coupling positive at the high scale. 
The lower values {\it i.e.} $\lambda_{H} < 0.18$ become negative at high scales 
and are ruled out, while $\lambda_{H} > 0.25$ is not obtained due to the chosen range of $M_{h_2}$ and upper bound on $\sin\theta$ coming from the Higgs signal strength measurements.
We have found that $\lambda_H$ has a very narrow range at the $m_t$ scale, which becomes broader at the pivot scale (horizon exit scale) after running the coupling, resulting in a wider range of $A_s$ at the pivot scale. Since we have considered the DM parameters in the feeble regime, this also places the quartic couplings in the feeble regime, except for the SM Higgs quartic coupling. Due to these feeble quartic couplings and to satisfy the Higgs inflation conditions where the inflaton field rolls along the SM Higgs direction, we have considered the dark Higgs non-minimal coupling $\xi_D = 0$ at the $m_t$ scale. 
However, $\xi_D$ is generated at a high scale due to the RG running effect of the parameters.
We have found many interesting correlations among the model parameters that control both inflation and DM physics. In particular, we obtained a linear correlation in the dark gauge coupling, $g_{D}$ and $\lambda_{HD}$ plane, where the allowed regions shrink significantly once we impose the upper bound of DM relic density on top of the inflation bounds.
For DM production, we have considered both decay and annihilation processes, incorporating new effects from gluon and photon annihilation channels mediated by loop diagrams at the DM production stage. All relevant vertices were implemented in 
\texttt{FeynRules} \cite{Alloul:2013bka} before being fed the 
\texttt{CalcHEP} \cite{Belyaev:2012qa} model files 
into \texttt{micrOMEGAs} \cite{Belanger:2018ccd}.
Finally, we have studied collider aspects in the $\kappa_3$-$\kappa_4$ plane
which are defined in section \ref{Collider-Prospects}. After incorporating both inflation and DM bounds,
we found that the $\kappa_{3,4}$ values deviate significantly from the SM 
values $(\kappa_{3}, \kappa_{4}) = (1,1)$ 
and the allowed region has promising detection prospects at the HL-LHC with 3 $ab^{-1}$ 
of data collection and within the allowed range from recent LHC data.

The rest of the paper is organised as follows: In section \ref{model}, 
we describe the model for the present study. In sections \ref{DM-physics} and 
\ref{inflation}, we briefly discuss DM and inflation, respectively. 
Section \ref{result} presents the results of our study. In Section \ref{Collider-Prospects}, we discuss the collider prospects of the allowed quartic couplings. Finally, in Section \ref{conclusion}, we provide our conclusions.

\section{Model}
\label{model}
The particle content of the present model is shown in Table \ref{tab:modelSM} 
and the associated Lagrangian can be expressed as,
\begin{center}
\begin{table}[t!]
\begin{center}
\begin{tabular}{||c|c|c|c||}
\hline
\hline
\begin{tabular}{c}
    Gauge\\
    Group\\ 
    \hline
    ${\rm SU(2)}_{\rm L}$\\  
    \hline
    ${\rm U(1)}_{\rm Y}$\\
        \hline
    ${\rm U(1)}_{\rm D}$\\  
\end{tabular}
&
\begin{tabular}{c|c|c}
    \multicolumn{3}{c}{Baryon Fields}\\ 
    \hline
    $Q_{L}^{i}=(u_{L}^{i},d_{L}^{i})^{T}$&$u_{R}^{i}$&$d_{R}^{i}$\\ 
    \hline
    $2$&$1$&$1$\\ 
    \hline
    $1/6$&$2/3$&$-1/3$\\
        \hline
    $0$&$0$&$0$\\  
\end{tabular}
&
\begin{tabular}{c|c}
    \multicolumn{2}{c}{Lepton Fields}\\
    \hline
    $L_{L}^{i}=(\nu_{L}^{i},e_{L}^{i})^{T}$ & $e_{R}^{i}$\\
    \hline
    $2$&$1$\\
    \hline
    $-1/2$&$-1$\\
        \hline
    $0$&$0$\\
\end{tabular}
&
\begin{tabular}{c|c}
    \multicolumn{2}{c}{Scalar Field}\\
    \hline
    $H$ & $\phi_{D}$\\
    \hline
    $2$ & $1$\\
    \hline
    $1/2$ & $0$\\
        \hline
    $0$ & $1$\\
\end{tabular}\\
\hline
\hline
\end{tabular}
\caption{Particle content of the present model and their corresponding charges under the SM and additional gauge groups.}
\label{tab:modelSM}
\end{center}
\end{table}
\end{center}

\begin{align}
    \mathcal{L} &= 
    \mathcal{L}_{{\rm SM}}
    - \frac{1}{4} W_{D\,\,\mu\nu} W_{D}^{\mu\nu}
    +|D\phi_D|^2 
    -V(\phi_D,H)\,.
    \label{total-lagrangian}
\end{align}
The first term $\mathcal{L}_{{\rm SM}}$ in Eq. \ref{total-lagrangian} corresponds to the Lagrangian 
for the SM sector, the second term represents the kinetic term for the extra $U(1)_{D}$
gauge boson, and finally, the last two terms represent the kinetic term for the extra scalar, and the potential consists of singlet and doublet scalars, respectively. In the kinetic term of the extra singlet scalar, the covariant derivative takes the form 
$D_{\mu} = \partial_{\mu} - i g_{D} W_{D\,\mu}$. The potential, up to dimension-4 terms, can be expressed as, 
\begin{align}
    V(\phi_D,H) = 
    - \mu^2_D \phi^{\dagger}_D \phi_D
    + \lambda_D (\phi^{\dagger}_D \phi_D)^2 
    - \mu^2_H H^{\dagger} H
    + \lambda_H (H^{\dagger} H)^2 
    + \lambda_{HD} \phi^{\dagger}_D \phi_D H^{\dagger} H
    \,.\label{eqn:scalar_pot_full}
\end{align}
Once the SM Higgs doublet and additional scalar develop spontaneous VEVs, the gauge symmetries break to a smaller group, we can write down the scalars in the Unitary gauge in the following way,
\begin{align}
    H =
    \frac{1}{\sqrt{2}}
    \begin{pmatrix}
    0 \\
    v_H + h  
    \end{pmatrix}
    \,,\quad
    \phi_D =
    \frac{v_D + \phi}{\sqrt{2}}
    \,.
\end{align}
The mass matrix for the scalars on the ($h\,\,\,\phi$) basis takes the following form,
\begin{align}
    M_{h\phi} = \begin{pmatrix}
   2 \lambda_H v_H^2 & \lambda_{HD} v_H v_D\\
    \lambda_{HD} v_H v_D & 2 \lambda_D v_D^2
    \end{pmatrix}
    \,,
\end{align}
We can diagonalise the above mass matrix and define the mass eigenbasis in terms of the flavour eigenbasis in the following way,
\begin{align}
    \begin{pmatrix}
    h_{1}\\
    h_{2}
    \end{pmatrix}
    = \begin{pmatrix}
    \cos\theta & -\sin\theta\\
    \sin\theta & \cos\theta
    \end{pmatrix}
    \begin{pmatrix}
    h\\
    \phi
    \end{pmatrix}\,.
    \label{higgs-mixing-matrix}
\end{align}

The quartic couplings, in terms of the masses of the two Higgses and the mixing angle, can be expressed as,
\begin{eqnarray}
\lambda_{H} &=& \frac{M^2_{h_2} + M^2_{h_1} - (M^2_{h_2} - M^2_{h_1}) \cos2\theta  }{4 v^{2}}\,,
\nonumber \\
\lambda_{D} &=& \frac{M^2_{h_2} + M^2_{h_1} + (M^2_{h_2} - M^2_{h_1}) \cos2\theta  }{4 v_{D}^{2}}\,,
\nonumber \\
\lambda_{HD} &=& \frac{(M^2_{h_2} - M^2_{h_1}) \cos\theta \sin\theta  }{v v_{D}}\,.
\nonumber \\
\label{quartic-coupling}
\end{eqnarray}
In the present work, we have considered $h_1$ as SM-like Higgs and $h_2$ as heavier BSM Higgs.
The constraints coming from perturbativity
and the requirement of the potential (in Eq.~\ref{eqn:scalar_pot_full}) is bounded from below, can be translated as,
\begin{eqnarray} 
0 \leq \lambda_{H,D, HD} \leq 4 \pi, \quad \lambda_{HD} + 2 \sqrt{\lambda_{H} \lambda_{D}} \geq 0. 
\label{pertubativity-bound}
\end{eqnarray}
Additionally, we cannot choose the mixing angle $\sin\theta$
to be arbitrarily large, as there is an upper bound on it, mainly coming from the recent precise measurements of the Higgs signal strength 
\cite{ATLAS:2022vkf, ATLAS:2016neq, Heo:2024cif},
which translates to the following upper bound on the mixing 
angle{}\footnote{For a detailed derivation, interested
readers are referred to Ref. \cite{Khan:2025yko}},
\begin{eqnarray} 
\sin\theta \leq 0.23,. 
\end{eqnarray}
Finally, when the $U(1)_{D}$ breaks down, then the extra gauge boson gets mass as follows,
\begin{align}
    M_{W_{D}} = g_D v_D
    \,.
\end{align}
The DM candidate can be made stable by assigning the charge conjugation symmetry in the dark sector as follows,
\begin{eqnarray}
\phi_{D} \rightarrow \phi^{\dagger}_{D}\,, W_{D\mu} \rightarrow - W_{D\mu}\,.
\end{eqnarray}
It is to be noted that by defining the above charge conjugation symmetry forbids the kinetic mixing with the hypercharge abelian gauge group. This explains the absence of the kinetic mixing term in the Lagrangian given in Eq.~\ref{total-lagrangian}. Moreover, when the dark $U(1)$ symmetry gets broken, the charge conjugation symmetry in the dark sector gets broken, and we have a remnant $\mathbb{Z}_2$ symmetry which ensures the DM stability. 
In studying the DM, we have considered the DM relic density in the range
as follows,
\begin{eqnarray}
10^{-5} \leq \Omega_{DM}h^{2} \leq 0.1226 
\label{DM-density-range}
\end{eqnarray}
The large range is considered 
to obtain more allowed points in a reasonable amount of time.
It is worth mentioning that our conclusion does not change if we consider the allowed $3\sigma$ range of the DM density from 
Planck data \cite{Planck:2018vyg}.
On the other hand, such a large range also allows us to easily
accommodate an additional DM candidate in our study,
which can account for the remaining components, similar to 
Refs. \cite{Costa:2022oaa, Belanger:2022gqc, Costa:2022lpy,
Khan:2023uii, Khan:2024biq}.

\section{Dark Matter}
\label{DM-physics}
The Boltzmann equation for the non-thermal DM evolution, considering both decay and annihilation contributions, can be expressed as,
\begin{eqnarray}
\frac{d Y_{W_{D}}}{dz} &=& \frac{2 M_{\text{pl}} z \sqrt{g_{*}(z)}}{1.66 M^2_{sc} g_{s}(z)}
\left[ \sum_{i = 1,2} \langle  \Gamma_{h_{i} \rightarrow W_{D}W_{D}} \rangle \left( 
Y^{\text{eq}}_{h_{i}} - Y^2_{W_{D}} \right) \right] \nonumber \\
&+& \frac{2 M_{sc} s(T)}{z^{2} H(T) T} \sum_{i = f, GB, h_{1,2}} \langle \sigma v \rangle_{ii \rightarrow W_{D}W_{D}} \left[ Y^{\text{eq}\,2}_{i}  - Y^2_{W_{D}} \right] 
\label{BE}
\end{eqnarray}
where $Y_{W_{D}}$ is the ratio of the number density of DM, $n_{W_{D}}$, and
entropy, $s(T)$, referred to as the co-moving number density.
The variable $z = \frac{M_{sc}}{T}$ where $M_{sc}$ is some reference scale and 
can be taken equal to the SM Higgs mass. The entropy $s(T)$ 
and the Hubble parameter $H(T) = \frac{\dot{a}}{a}$ (where $a$ is scale
factor) can be expressed as,
\begin{eqnarray}
s(T) =  \frac{2 \pi^{2}}{45} g_{s} T^{3}\,,\,\,\, 
H(T) = \sqrt{\frac{8 \pi^{3} g_{\rho}}{90}} \, \frac{T^{2}}{M_{\text{pl}}}\,. 
\end{eqnarray}
where $g_{s}(T)$ is the effective {\it d.o.f} of entropy,  $g_{\rho}(T)$  is effective relativistic {\it d.o.f} of energy energy density of the Universe at temperature $T$. The factor $g_{*}(T)$ is defined as,
\begin{align}
    g_{*}^{1/2}(T) &= \frac{g_s(T)}{g_{\rho}(T)} \left( 1 + \frac{ T}{3 g_{s}(T)} \frac{dg_s(T)}{dT}\right)
\end{align}

The thermal average of the decay width ($\Gamma_{h_{i} \rightarrow W_{D}W_{D}}$) and the annihilation cross-section ($\sigma_{ii \rightarrow W_{D} W_{D}}$) times velocity can be
expressed as \cite{Gondolo:1990dk},

\begin{eqnarray}
&& \langle \Gamma_{h_{i} \rightarrow W_{D}W_{D}} \rangle = 
\Gamma_{h_{i} \rightarrow W_{D}W_{D}}\, \frac{K_{1}\left( \frac{m_{h_i}}{M_{sc}}z \right)}
{K_{2}\left( \frac{m_{h_{i}}}{M_{sc}}z \right)}, \nonumber \\
&& \langle \sigma v \rangle_{ii} = \frac{g^2_{i} T}{2 \left( 2 \pi \right)^{4} 
\left(n^{{\text{eq}}}_{i}\right)^{2} } \int^{\infty}_{4M_{i}^2}
\sigma_{ii \rightarrow W_{D} W_{D}} (s - 4 M^2_{i}) \sqrt{s} K_{1} 
\left(\frac{\sqrt{s}}{T} \right)\,ds ,  
\end{eqnarray}
where $n_{i}^{\text{eq}} = g_{i} \frac{M_{i}^2}{2\pi^2} T K_2(\frac{M_i}{T}) $ 
is the number density of the species $i = l, q, W,Z, \gamma,g, h_{1,2}$
with $g_i$ and $M_i$ being the {\it d.o.f} and mass of the $i^{th}$ particle. We have 
included all the particle spectra in our analysis, including both decay and
annihilation. We have produced our DM results using the micrOMEGAs
which is publicly available code for solving the Boltzmann equations and for the CalcHEP file generation, we have
used the FeynRules package. Moreover, we have included the one-loop
contribution to the DM production from the gluon and photon annihilation 
through a loop process mediated by the fermionic particles. 
Once we determine the co-moving number density for the DM, we can determine the DM relic density using the following expression \cite{Edsjo:1997bg},
\begin{eqnarray}
\Omega_{W_{D}} h^{2} = 2.755 \times 10^{8} \left( \frac{M_{W_{D}}}{\rm GeV} 
\right) Y_{W_{D}}\,.
\label{dm-density-expression}
\end{eqnarray} 
\begin{figure}[t!]
    \centering
    \includegraphics[angle=0,height=6.5cm,width=10.5cm]{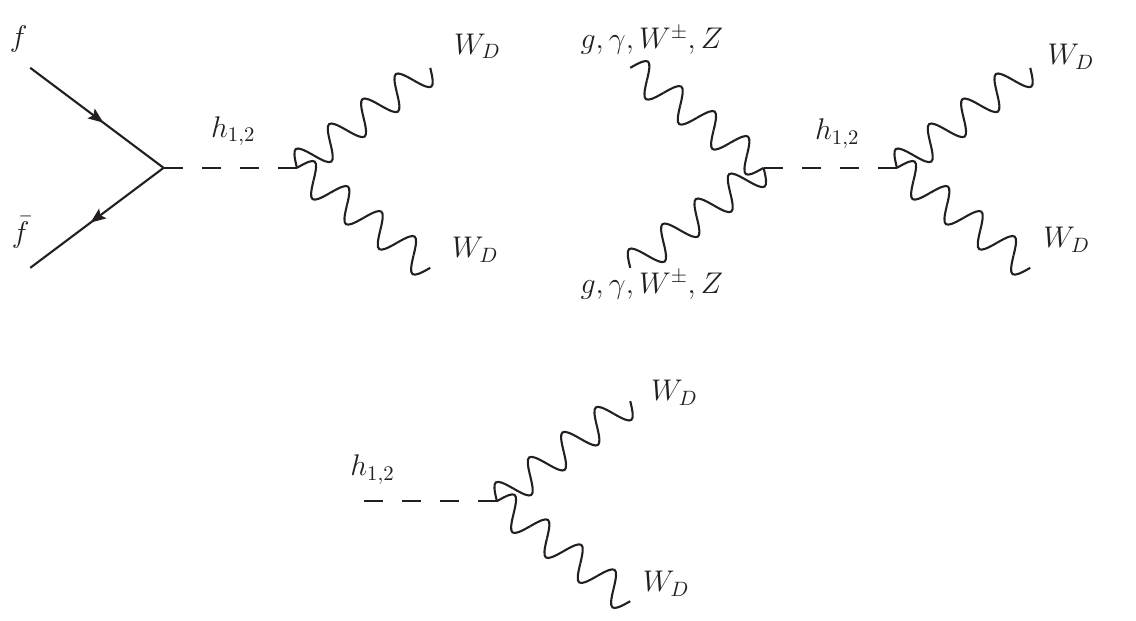}
    \caption{Feynman diagrams relevant for the production of the vector DM in the
    present study.}
    \label{feynman-diagram}
\end{figure}

The relevant diagrams for the vector DM production from the decay and annihilation
of scalars, fermions and gauge bosons are shown in Fig. \ref{feynman-diagram}. 
We can calculate the DM production analytically from the gauge boson 
annihilation diagrams, although our numerical results are presented based on the analysis using micrOMEGAs. In this part, we display the DM production from gluon 
annihilation by following the Ref. \cite{Hall:2009bx}. The Boltzmann equation in Eq. \ref{BE} can be written as follows by using the temperature as a variable instead of $z = \frac{M_{sc}}{T}$ and considering only the annihilation contribution,
\begin{eqnarray}
\frac{d Y^{gg}_{W_D}}{d T} = - \frac{1}{37.77 \pi^{7} g_{s}(T) \sqrt{g_{*}(T)}\, T^{5} }
\int^{\infty}_{0} ds \sqrt{s}\, |M|^{2}\, K_{1}\left( \frac{\sqrt{s}}{T} \right) 
\label{BE-integration}
\end{eqnarray}
where the amplitude for the $gg \rightarrow W_{D}W_{D}$ can be expressed as,
\begin{eqnarray}
|M|^{2} &=& \frac{1}{4 N^2_{c}} \left| \frac{g_{h_{i} W_{D}W_{D}} g_{h_{i} gg} }
{s-M^2_{h_i}} \right|^{2} \,s^2 \left[ 1 + \frac{(s-2 M^2_{W_D})^{2}}{8 M^4_{W_D}} \right] \nonumber \\
&\xRightarrow[]{s \gg }& \frac{1}{32 N^2_{c}} 
\biggl| \sum_{i} g_{h_{i} W_{D}W_{D}} g_{h_{i} gg}  \biggr|^{2} \frac{s^{2}}
{M^4_{W_D}} 
\end{eqnarray} 
where $N_c$ is the colour charge for the gluon. Finally substituting the $|M|^2$
for very high value of the $s$ in Eq. \ref{BE-integration} and doing the $s$ integration and temperature integration from early temperature $T_{\text{ini}}$ to some lower temperature 
$T_0$, we get the co-moving number density as follows,
\begin{eqnarray}
Y^{gg}_{W_D} = \frac{2.244 \times 10^{-3}}{g_{s}(T) \sqrt{g_{*}(T)}}
\left( T^3_{\text{ini}} - T^3_{0} \right) \biggl( \frac{ \left| \sum_{i=1,2} 
g_{h_{i} W_{D} W_{D} } g_{h_{i} gg} \right|^{2} }{32 N^2_{c} M^4_{W_D}} \biggr),
\label{co-moving-WD}
\end{eqnarray}
the couplings $g_{h_{i}W_{D}W_{D}}$ and $g_{h_{i}gg}$ can be expressed as
follows,
\begin{eqnarray}
&& g_{h_{1}W_{D}W_{D}} = -2 \sin\theta g_{D} M_{W_{D}}\,, 
g_{h_{2}W_{D}W_{D}} = 2 \cos\theta g_{D} M_{W_{D}} \nonumber \\
&& g_{h_{1} gg} = \frac{\alpha_S \cos\theta }{4 \pi v}\,
\kappa\, f\left(\frac{4 M^2_{t}}{M^2_{h_{1}}}\right),\,
g_{h_{2} gg} = \frac{\alpha_S \sin\theta }{4 \pi v}\,
\kappa\, f\left(\frac{4 M^2_{t}}{M^2_{h_{2}}}\right),
\label{gluon-vertices}
\end{eqnarray}
where $\alpha_S = 0.1184$, $N_c = 8$ for gluon, $M_t = 172.5$ GeV is the top
quark mass at the Z-pole mass and 
$\kappa^2 = 3$ coming from the higher 
loop corrections \cite{Catani:2001ic, Harlander:2002wh} 
referred to as the K-factor in literature. 
The function $f(x) = - 2 x \biggl( 1 + (1-x) g(x) \biggr)$ 
coming from the one-loop diagram
and $g(x)$ can be expresses as \cite{Djouadi:2005gi},
\begin{eqnarray}
g(x) &=& arcsin\biggl( \frac{1}{\sqrt{x}} \biggr)\,\,{\rm for}\,\,x \ge 1 \nonumber 
\nonumber \\
&=& -\frac{1}{4} \biggl[ \log\biggl( \frac{1 + \sqrt{1-x}}{1 - \sqrt{1-x}} \biggr)
- i \pi \biggr]^{2}\,\,{\rm for}\,\,x < 1\,,.
\end{eqnarray} 
where $x = \frac{4 M^2_{t}}{M^2_{h_{1,2}}}$ is coming from the top loop contribution which
is dominant over the other loop-mediating particles.
In the co-moving number density in Eq. \ref{co-moving-WD}, we can take $T_{0} = 0$, which will 
not change the final output, and we can determine
the DM density using Eq. \ref{dm-density-expression} for the $gg$ channel. 
From the co-moving expression, we can see that DM production from gluon annihilation is dependent on the starting temperature $T_{\text{ini}}$. It is to be noted that we get the same expression for the photon annihilation 
case as well, except we have $N_{c} = 1$ for the photon instead of $N_{c} = 8$ for the gluon. 
All these vertices of gluon and photon with Higgses in Eq. \ref{gluon-vertices} have been implemented in the \texttt{Feynrules} for the DM study using the \texttt{micrOMEGAs} package.

\section{Inflation}
\label{inflation}
After thoroughly discussing the DM production mechanism in the present model, we now turn to the possible realisation of inflation within the same framework.
The relevant action, in unitary gauge, can be expressed as,
\begin{align}
    S = \int d^4x \, \sqrt{-g_{\rm J}} \, \left[
    \frac{M_{\rm P}^2}{2}\left(
    1 + \xi_H \frac{h^2}{M_{\rm P}^2} 
    + \xi_D \frac{\phi^2}{M_{\rm P}^2}
    \right)R_{\rm J}
    +\frac{1}{2}g_{\rm J}^{\mu\nu}\partial_\mu h \partial_\nu h
    +\frac{1}{2}g_{\rm J}^{\mu\nu}\partial_\mu \phi \partial_\nu \phi
    -V(\phi,h)
    \right]
    \,,\label{eqn:inflation_action_jordan}
\end{align}
where we use the subscript J to indicate the Jordan frame. The nonminimal couplings $\xi_H h^2 R$ and $\xi_D \phi^2 R$ naturally appear due to their mass dimension of four and are generated via radiative corrections even if initially set to zero. Here, we focus on positive nonminimal couplings.
The tree-level scalar potential is
\begin{align}
V(\phi,h) = \frac{1}{4}\lambda_H h^4 + \frac{1}{4}\lambda_D \phi^4 + \frac{1}{4}\lambda_{HD}\phi^2h^2\,,
\label{eqn:inflaton_pot_jordan}
\end{align}
where quadratic mass terms are neglected during inflation. The model reduces to standard Higgs inflation \cite{Bezrukov:2007ep} in the $\phi\rightarrow 0$ limit.
Inflation with the Higgs-portal coupling is well studied, {\it e.g.}, in Ref.~\cite{Lebedev:2011aq}, which we closely follow. We first discuss classical inflation and derive the relevant constraints for the Higgs inflation. After this, we move to quantum analysis of inflation, incorporating RG running of the couplings. The detailed calculation in obtaining the relevant action has been provided in Ref. \cite{Khan:2023uii}.

\subsection{Classical analysis}
\label{subsec:InfClassical}
The Jordan-frame action as shown in Eq. \ref{eqn:inflation_action_jordan} can be transformed to the Einstein frame, denoted by the subscript E, via Weyl rescaling\footnote{We have followed the metric convention $g_{\mu\nu} = \left(+\,-\,-\,- \right).$},
\begin{align}
    g_{{\rm J} \mu\nu} \rightarrow 
    g_{{\rm E} \mu\nu} = \Omega^2 g_{{\rm J} \mu\nu}
    \,,
\end{align}
with the conformal factor
\begin{align}
    \Omega^2 = 1 + \xi_H \frac{h^2}{M_{\rm P}^2}
    + \xi_D \frac{\phi^2}{M_{\rm P}^2}
    \,.
\end{align}
After the Weyl transformation, the action shown in Eq. \ref{eqn:inflation_action_jordan} takes the 
following form in the Einstein frame,
\begin{align}
    S = \int d^4x \, \sqrt{-g_{\rm E}} \, &\bigg[
    \frac{M_{\rm P}^2}{2}R_{\rm E}
    +\frac{3}{4}M_{\rm P}^2g_{\rm E}^{\mu\nu}\partial_\mu\ln\Omega^2\partial_\nu\ln\Omega^2
    \nonumber\\
    & 
    +\frac{1}{2\Omega^2}g_{\rm E}^{\mu\nu}\partial_\mu h \partial_\nu h
    +\frac{1}{2\Omega^2}g_{\rm E}^{\mu\nu}\partial_\mu \phi \partial_\nu \phi
    -\frac{V}{\Omega^4}
    \bigg]\,.
\end{align}
After defining the fields $  \varphi \equiv \sqrt{\frac{3}{2}}M_{\rm P}\ln\Omega^2\,$ and $ \chi \equiv \frac{\phi}{h}$ and in the large field limit where
    $\Omega^{2} \gg 1$, we get the action as \cite{Khan:2023uii}, 
\begin{align}
    S = \int d^4x \, \sqrt{-g_{\rm E}} \, &\bigg[
    \frac{M_{\rm P}^2}{2}R_{\rm E}
    +\frac{1}{2} g_{\rm E}^{\mu\nu}\partial_\mu \varphi \partial_\nu \varphi 
    +\frac{1}{2} g_{\rm E}^{\mu\nu}\partial_\mu \chi_c \partial_\nu \chi_c 
    \nonumber\\
    &\qquad
    +g_{\rm E}^{\mu\nu}\partial_\mu \varphi \partial_\nu \chi_c
    \frac{(\xi_H-\xi_D)\chi}{\sqrt{6}\sqrt{\xi_H^2+\xi_D^2\chi^2}\sqrt{\xi_H+\xi_D\chi^2}}
    -U
    \bigg]\,,\label{eqn:effective-action-inflation}
\end{align}    
where the field $\chi$ has also been 
canonically normalised via,
\begin{align}
    \left(\frac{d\chi_c}{d\chi}\right)^2 =
    \frac{M_{\rm P}^2(\xi_H^2+\xi_D^2\chi^2)}{(\xi_H+\xi_D\chi^2)^3}
    \,.
\end{align}
The potential $U$ is in the Einstein frame, which takes the following form,     
     \begin{align}
    U = \frac{V}{\Omega^4} = \frac{\lambda_H + \lambda_{HD}\chi^2 + \lambda_D\chi^4}{4(\xi_H+\xi_D\chi^2)^2}
    \left(
    1 - e^{-\sqrt{\frac{2}{3}}\frac{\varphi}{M_{\rm P}}}
    \right)^2
    M_{\rm P}^4
    \,.
\end{align}   
For a finite, nonzero $\chi_c$, the kinetic mixing term in 
Eq. \ref{eqn:effective-action-inflation} disappears when $\xi_H=\xi_D$. 
On the other hand, when $\xi_H \neq \xi_D$, 
this term becomes suppressed in the presence of a large nonminimal coupling.
Our primary focus is on scenarios where at least one of the nonminimal couplings (in particular $\xi_H$) is sufficiently large to allow the kinetic mixing term to be safely neglected. 
For other choices of parameterisation, one needs to diagonalise the kinetic terms and introduce new fields in the diagonal basis.
Inflation can proceed along the SM Higgs direction, the dark Higgs direction, or a combination of both. We refer to these cases as SM Higgs inflation, dark Higgs inflation, and mixed inflation, respectively. 
In Ref. \cite{Khan:2023uii}, all three cases have been discussed in detail and
in the present work, we explore only the Higgs inflation case and how it gets constrained
from the freeze-in DM. 

{\bf SM Higgs inflation scenario}: 

The SM Higgs inflation corresponds to the $\chi=0$ case, and we need
to maximise the potential with the first and second-order derivatives,
\begin{align}
    \frac{\partial U}{\partial \chi_c}\bigg\vert_{\chi=0} &= 0
    \,,\\
    \frac{\partial^2 U}{\partial \chi_c^2}\bigg\vert_{\chi=0} &=
    \frac{M^2(\lambda_{HD}\xi_H - 2\lambda_H\xi_D)}{2\xi_H^2}\left(
    1 - e^{-\sqrt{\frac{2}{3}}\frac{\varphi}{M_{\rm P}}}
    \right)^2\,.
\end{align}
Thus, $\chi=0$ becomes minimum only when $\frac{\partial^2 U}{\partial \chi_c^2} > 0$ which implies,
\begin{align}
\lambda_{HD} - 2\lambda_H\frac{\xi_D}{\xi_H} > 0\,.
\label{eqn:SMHIconditiondv}
\end{align}
After satisfying the Eq. \ref{eqn:SMHIconditiondv}, we can work with the action,
\begin{align}
    S = \int d^4x \, \sqrt{-g_{\rm E}} \, &\bigg[
    \frac{M_{\rm P}^2}{2}R_{\rm E}
    +\frac{1}{2} g_{\rm E}^{\mu\nu}\partial_\mu \varphi \partial_\nu \varphi 
    -\frac{M_{\rm P}^4 \lambda_H}{4\xi_H^2}\left(
    1 - e^{-\sqrt{\frac{2}{3}}\frac{\varphi}{M_{\rm P}}}
    \right)^2
    \bigg]\,,
\end{align}
which is similar to the standard nonminimally coupled single-field model,
allowing us to determine inflationary observables such as the spectral 
index $n_s$, the tensor-to-scalar ratio $r$ and curvature power spectrum amplitude $A_s$. 
Inflationary observables can be expressed in terms
of the slow-roll parameters, which are defined as follows,
\begin{align}
    \epsilon = \frac{M_{\rm P}^2}{2}\left(
    \frac{U'}{U}
    \right)^2
    \,,\quad
    \eta = M_{\rm P}^2 \frac{U''}{U}
    \,,\quad 
    \kappa^2 = M_{\rm P}^4\frac{U'U'''}{U^2}
    \,,
\end{align}
where $U' = \frac{d U}{d \varphi}$ and same for higher derivative.
The $n_s$, $r$, and the curvature power spectrum amplitude $A_s$ in terms of the slow roll 
parameters can be expressed as \cite{Stewart:1993bc,Liddle:1994dx,Leach:2002ar},
\begin{align}
    n_s &=
    1 - 6\epsilon + 2\eta 
    - \frac{2}{3}(5+36c)\epsilon^2
    + 2(8c-1)\epsilon\eta + \frac{2}{3}\eta^2
    +\left(\frac{2}{3}-2c\right)\kappa^2
    \,,\label{eqn:SSI}\\
    r &=
    16\epsilon\left[
    1 + \left(
    4c-\frac{4}{3}
    \right)\epsilon + \left(
    \frac{2}{3} - 2c
    \right)\eta
    \right]
    \,,\label{eqn:TTSR} \\
     A_s &= \frac{U}{24\pi^2M_{\rm P}^4\epsilon}\,
     \label{eqn:PSamp}
\end{align}
where $c=\gamma + \ln 2 - 2$ with $\gamma \approx 0.5772$. 
The aforementioned quantities need to be evaluated at the horizon exit scale at the beginning of the inflation.
The number of $e$-folds can be computed as,
\begin{align}\label{eqn:efolds}
    N = -\frac{1}{M_{\rm P}^2}\int_{\varphi_*}^{\varphi_e} \frac{U}{U'} d\varphi \,,
\end{align}
where $\varphi_e$  and $\varphi_*$ denote the scale of the end of inflation and the horizon exit, respectively.
In general $\frac{\ddot{a}}{a} = \left(1 -\epsilon \right) H^{2}$ and the end of the inflation scale,$\varphi_e$, implies 
when $\epsilon \simeq 1$. Once we have $\varphi_e$ then we can evaluate 
the integral in Eq. \ref{eqn:efolds}, to find the value of the 
 horizon exit scale $\varphi_*$, until we get
the integral value $N \simeq 60$.   

\subsection{Quantum analysis using RG running}
\label{subsec:InfQuantum}

This part considers the inflation study using the renormalisation group
(RG) equations from the DM observables to the inflation observables when
the horizon exit happens. 
Following the procedures outlined in Ref.~\cite{Kim:2014kok, Khan:2023uii}, 
we consider the RG-improved effective action in the Jordan frame.  As shown in Sec. \ref{subsec:InfClassical}, the relevant inflationary action can be effectively represented by a single-field action.  Therefore, we consider the following leading effective action for the Higgs field only,
\begin{align}
    \Gamma_{\rm eff} = \int d^4x \, \sqrt{-g_{\rm J}}\left[
    \frac{M_{\rm P}^2}{2} \Omega^2(t) R_{\rm J}
    +\frac{1}{2}g^{\mu\nu}_{\rm J}G^2(t)\partial_\mu h(t)\partial_\nu h(t)
    -V_{\rm eff}(t)
    \right]\,,
\end{align}
where $t=\ln(\mu/M_t)$, $\mu$ is the renormalisation scale and 
$M_t$ is the top-quark pole mass. The other quantities are defined as,
\begin{align}
    \Omega^2(t) &= 1 + \xi_H(t) G^2(t) \frac{h^2(t)}{M_{\rm P}^2}
    \,,\\
    V_{\rm eff}(t) &= \frac{\lambda_H(t)}{4}G^4(t)h^4(t)
    \,,\\
    G(t) &= \exp\left(
    -\int^t dt' \, \frac{\gamma_h}{1+\gamma_h}
    \right)\,,
\end{align}
The RG equations for all the parameters as well as the anomalous dimensions are presented in Appendix~\ref{apdx:RGEs}.
In the Einstein frame, the effective action takes the form,
\begin{align}
    \Gamma_{\rm eff} = \int d^4x \, \sqrt{-g_{\rm E}}\left[
    \frac{M_{\rm P}^2}{2}R_{\rm E}
    +\frac{1}{2}g_{\rm E}^{\mu\nu}\partial_\mu\Psi(t)\partial_\nu\Psi(t)
    -U_{\rm eff}(t)
    \right]\,,
\end{align}
where $\Psi$ is defined canonically normalised way,
\begin{align}
    \left(
    \frac{\partial \Psi}{\partial h}
    \right)^2 =
    \frac{G^2}{\Omega^2}
    +\frac{3M_{\rm P}^2}{2\Omega^4}\left(
    \frac{d\Omega^2}{d h}
    \right)^2
    \,,
\end{align}
and effective potential in the Einstein-frame is
\begin{align}\label{eqn:EFeffPot}
    U_{\rm eff}(t) =
    \frac{\lambda_H(t)G^4(t)h^4(t)}{4\biggl(1+\xi_H(t)G^2(t)h^2(t)/M_{\rm P}^2\biggr)^2}\,.
\end{align}

In the present work, we satisfy the inflation condition 
as given by Eq. \ref{eqn:SMHIconditiondv} at the inflation scale
after using RG equations. The detailed description of finding 
the inflation scale are presented in Refs. \cite{Kim:2014kok}. 
The PDG values of the
EW parameters \cite{ParticleDataGroup:2022pth} at the top-quark pole mass $M_t$ are $M_t=172.5$ GeV, $M_W=80.377$ GeV, $M_Z=91.1876$ GeV, $M_{h_1}=125.25$ GeV, $G_F=1.1664\times 10^{-5}$ GeV${}^{-2}$, $\alpha=1/127.951$, and $\alpha_s=0.118$, where $G_F, \alpha$ and $\alpha_s$ are the Fermi constant, 
fine-structure constant and strong coupling constant at $M_Z$ 
\cite{Buttazzo:2013uya}.
After setting the EW parameters at the aforementioned values, we run the model parameters
$\{\lambda_H, \lambda_{HD}, \lambda_D, g_D, \xi_H\}$ from
 $\mu=M_t$ to Planck scale $\mu=M_{\text{pl}}$ using the RG equations shown 
 in Appendix~\ref{apdx:RGEs}.
 We choose $\xi_D = 0$ at top pole mass\footnote{Although it is generated at the inflation scale through the RG running.}, which is necessary; otherwise, we cannot satisfy the Higgs inflation condition in Eq. \ref{eqn:SMHIconditiondv} because $\lambda_{HD}$ lies in the feeble regime due to the choice of the feeble gauge coupling $g_D$ for the FIMP DM study. In the present work, the Higgs quartic coupling gets an additional contribution from the second Higgs mass and Higgses mixing angle which increases its value from the SM predicted value. This increment  
  makes the Higgs potential safe during the RG running and avoids the 
  instability problem of the SM Higgs potential up to the Planck scale. As said before, we find the end of the inflation scale by setting 
  $\epsilon \simeq 1$ and the horizon exit scale by imposing the 60 $e-$folds.
Once we determine the horizon exit, we then examine the inflation
condition as given in Eq. \ref{eqn:SMHIconditiondv} and the perturbativity together
with the instability condition, $\lambda_{H,D,HD} > 0$\,.
The SM Higgs non-minimal coupling has been chosen in such a way so that we get the amplitude of the curvature power spectrum, defined in Eq. \ref{eqn:PSamp}, as $A_s \simeq 2.1 \times 10^{-9}$ at the pivot scale with slight variation around this value to get more points.
Planck collaboration has measured $A_s$, $n_s$, and  $r$, 
reporting their values with 68\% confidence intervals as follows \cite{Planck:2018vyg},
\begin{eqnarray}
A_s = \left(2.105 \pm 0.030 \right)\times 10^{-9}, 
n_s = 0.9665 \pm 0.0038\,,
\,\,{\rm and}\,\,r < 0.106\,. 
\label{As-ns-r-label}
\end{eqnarray} 

\subsection{Validity of the Higgs Inflation}

In Ref. \cite{Bezrukov:2007ep}, Higgs inflation was first explained using the 
non-minimal coupling of the SM Higgs with
the Ricci scalar, similar to the action shown 
in Eq. \ref{eqn:inflation_action_jordan}, 
but with the SM Higgs only.
In a subsequent study \cite{Bezrukov:2008ej}, 
based on the same setup, the authors studied the validity of Higgs
inflation using the RG-running of the couplings. In Ref. \cite{Bezrukov:2008ut}, they also addressed
reheating in the same setup and the successful generation of the SM bath. Later on, many studies
questioned the Higgs as an inflaton based on the unitarity violation beyond the cutoff scale
of the effective theory. For example, in Ref. \cite{Burgess:2009ea}, 
the authors pointed out the unitarity
issue based on the power counting formalism of effective field theory. A similar outcome has
also been pointed out in Ref. \cite{Barbon:2009ya}, where they claimed that the naturalness of the
proposed Higgs inflation is lost because the effective theory appears to cease to be valid
beyond the cutoff scale $\Lambda = \frac{M_{\text{pl}}}{\xi}$, whereas Higgs inflation
requires the field value $h > \frac{M_{\text{pl}}}{\sqrt{\xi}} > \frac{M_{\text{pl}}}{\xi}$. Therefore,
it is not recommended to extrapolate the theory beyond the cutoff without knowing the full
microscopic theory. In Refs. \cite{Lerner:2009na, Burgess:2010zq, Hertzberg:2010dc}, 
similar conclusions have been drawn regarding the
issues of Higgs inflation and unitarity. In Refs. \cite{Bezrukov:2010jz}, 
it has been shown that the cutoff scale
of the effective theory is background field dependent, and during the study of inflation, the cutoff
is always above the field value. Therefore, the above critical issues are not directly
applicable to Higgs inflation using the non-minimal coupling. Therefore, even though the theory is not UV complete and always has a cutoff, inflation can be studied safely because the cutoff is background field dependent. Finally, the main problem with SM Higgs inflation
arises due to the fixed value of the Higgs quartic coupling $\lambda_{H} = 0.125$, which demands a high value
of the non-minimal coupling $\xi$ to explain the amplitude of the curvature power spectrum obtained from the CMB. These issues will not arise when we consider an extra singlet
as the inflaton because no such restriction exists for the dark Higgs quartic coupling,
and inflation can be achieved for smaller values of $\xi$, which is left for future
study in the context of freeze-in DM.

\section{Results}
\label{result}
In this section, we present our results after scanning the model parameters in
the following range at the top quark pole mass,
\begin{eqnarray}
&&10^{-14} \leq g_{D} \leq 10^{-9}\,,\,\, 1 \leq M_{W_{D}}\,[GeV] \leq 10^{3}\,,
10^{-3} \leq \sin\theta \leq 0.2\,,\nonumber \\
&& 1 \leq (M_{h_{2}} - M_{h_{1}})\,\,[GeV] \leq 10^{3}\,,
10^{4} \leq \xi_{H} \leq 1.5 \times 10^{4}, \xi_{D} = 0
\end{eqnarray}

In the above range of parameters, we have considered the gauge coupling value in the feeble regime and the dark non-minimal coupling to zero. Therefore, we can solve the 
differential equation by assuming that $\xi_D$ is negligible\footnote{This is a valid assumption because using the running of the non-minimal coupling
we always get smaller value of $\xi_D$ upto Planck scale.} as follows,
\begin{eqnarray}
\frac{1}{g^2_{D}(\mu)} = \frac{1}{g^2_{D}(\mu_0)} - \frac{1}{6 \pi^2} 
\ln \left( \frac{\mu}{\mu_0} \right)\,.
\end{eqnarray}
where $\mu$ is some higher scale and $\mu_0$ is the top-quark pole mass and 
$g_{D}(\mu)$ and $g_{D}(\mu_0))$ are defined at those scales, respectively.
From the above equation, we can see that 
$\frac{1}{g^2_{D}(\mu_0)} \gg 
\frac{1}{6 \pi^2} 
\ln \left( \frac{\mu}{\mu_0} \right)$, due to the choice of the feeble gauge coupling,
so in our study we have $g_{D}(\mu) \simeq g_{D}(\mu_0)$.

\begin{figure}[h!]
\centering
\includegraphics[angle=0,height=7.5cm,width=7.5cm]{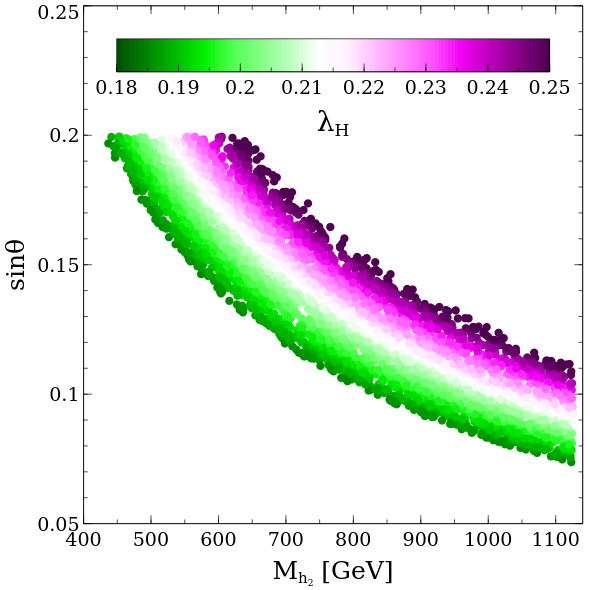}
\includegraphics[angle=0,height=7.5cm,width=7.5cm]{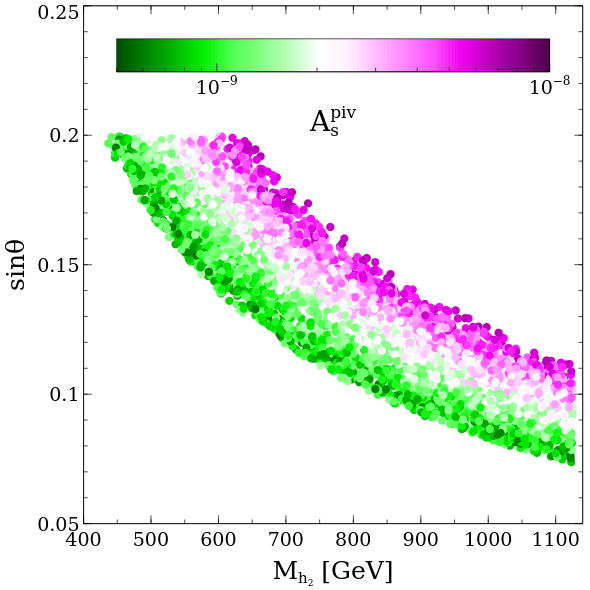}
\includegraphics[angle=0,height=7.5cm,width=7.5cm]{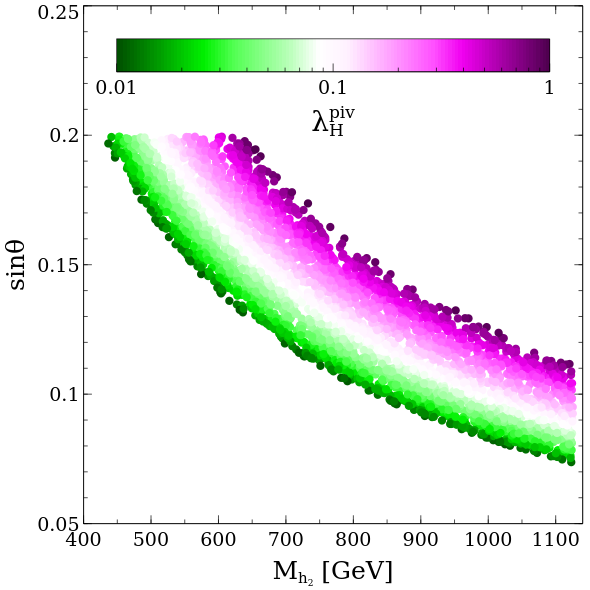}
\includegraphics[angle=0,height=7.5cm,width=7.5cm]{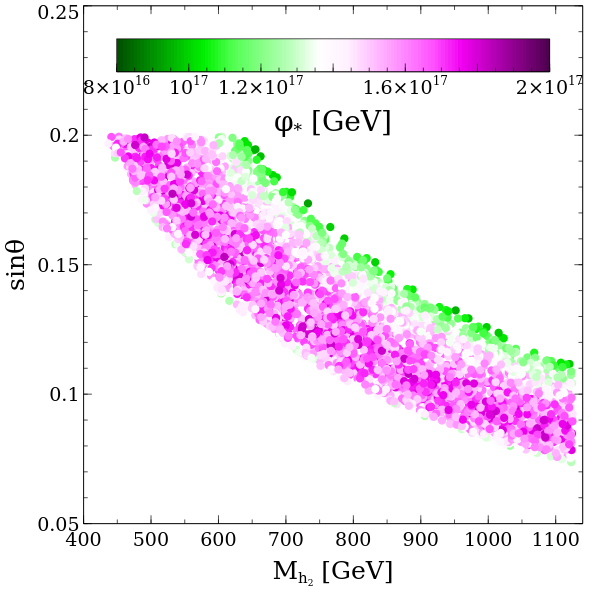}
\caption{Upper LP shows the scatter plot in the $M_{h_2} - \sin\theta$ plane, where 
the colour bar shows the SM Higgs quartic coupling. 
Upper RP also shows the scatter plot in the same plane, but the colour bar shows
the variation in the amplitude of the curvature power 
spectrum at the horizon exit scale denoted as $A^{\text{piv}}_{s}$. In the lower
LP, we have shown the same plane but the colour bar shows a different
values of the Higgs quartic coupling at the pivot scale after running 
$\lambda_{H}$ up to the pivot scale whereas the RP shows the 
value of the pivot scale $\phi_{*}$, respectively.} 
\label{scatter-plot-1}
\end{figure}

In Fig. \ref{scatter-plot-1}, we have shown scatter plots in the 
$M_{h_2}-\sin\theta$ plane and
the colour variations represent the different values of quartic coupling $\lambda_{H}$
and the amplitude of the curvature power spectrum at pivot scale ($A_s^{\text{piv}}$) in the upper LP and RP, respectively.
In both the plots, the data points have been obtained after
running all the couplings 
upto the Planck scale and satisfied the inflation conditions described in 
section \ref{inflation} as well as the perturbativity bounds and 
the quartic couplings are always positive. In general, the quartic coupling 
$\lambda_{H} \propto M^2_{h_{2}} \sin^{2}\theta$ as shown in 
Eq. (\ref{quartic-coupling}) which is also visible from the colour variation.
From the plot, we can see that $\lambda_{H} < 0.18$ is not allowed because those
values go negative at the high scale when we run all the parameters upto the
Planck scale which violates the inflation conditions. 
On the other hand, $\lambda_{H} > 0.25$ is also not allowed because
of the bound on the mixing angle from collider $\sin\theta < 0.23$ and the choice 
of $M_{h_{2}}$ mass range. 
In the RP, we have shown the scatter plot in the same plane 
but the colour variation shows the different values of the 
amplitude of the curvature 
power spectrum $A^{\text{piv}}_s$ at the horizon exit scale. 
We can see from the plot that the region 
which corresponds to larger values of $\lambda_{H}$ represents the higher values 
of the $A_s^{\text{piv}}$. It is worth to mention that $A^{\text{piv}}_{s}
 \propto \frac{\lambda_{H}}{\xi^2_{H}\, \epsilon_{\text{piv}}} $, therefore, the
 value increases with the increment of $\lambda_H$ at the pivot scale.  
The large variation in $A_s^{\text{piv}}$ will be clear from the lower panel plots, 
where two plots have been displayed 
in the same plane but the colour variation represents the different parameters.
In the lower LP, we have shown the Higgs quartic coupling at the pivot scale which
can differ by $\mathcal{O}(10^{2})$ contrary to the variation of the Higgs
quartic coupling at the top quark pole mass which differs at 
most $\mathcal{O}(0.07)$.
In the lower RP, we have shown the variation in the colour bar for different values
of the pivot scale, which is basically the horizon exit at the early Universe
and how it changes with the values of Higgs mass $M_{h_2}$ and 
mixing angle $\sin\theta$. We can see that the variation is not significant 
in the range of the parameters we have considered. A slight variation in $\varphi_{*}$ arises due to the RG running of $\xi_H$. At higher energy scales, $\xi_H$ depends on the Higgs quartic coupling $\lambda_H$, which changes with different values of $\sin\theta$ and $M_{h_2}$.
Therefore, the variation in 
$A^{\text{piv}}_{S}$ mainly comes due to the variation in $\lambda^{\text{piv}}_{H}$
with subdominant diminishing effect comes from the $\xi_H$ and $\phi_{*}$.

\begin{figure}[h!]
\centering
\includegraphics[angle=0,height=7.5cm,width=7.5cm]{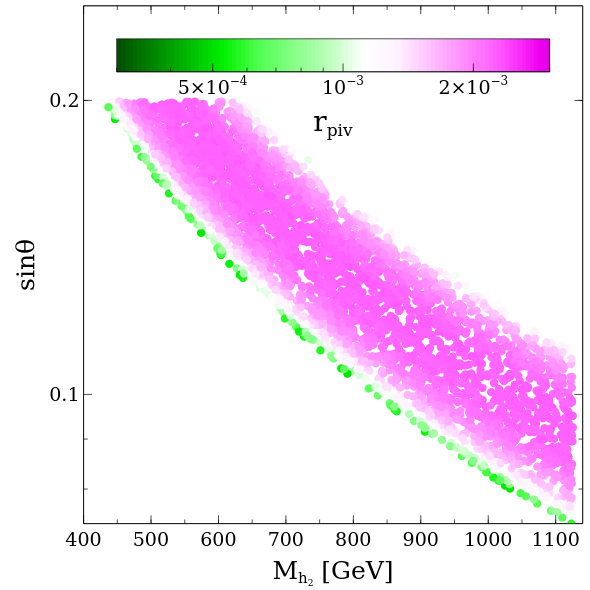}
\includegraphics[angle=0,height=7.5cm,width=7.5cm]{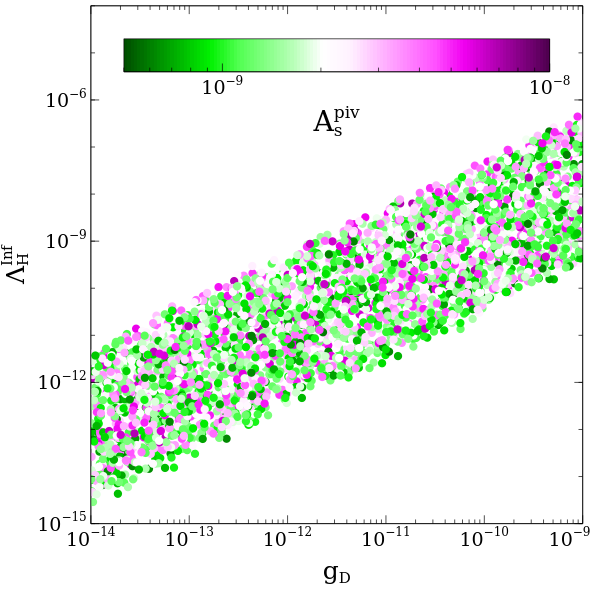}
\includegraphics[angle=0,height=7.5cm,width=7.5cm]{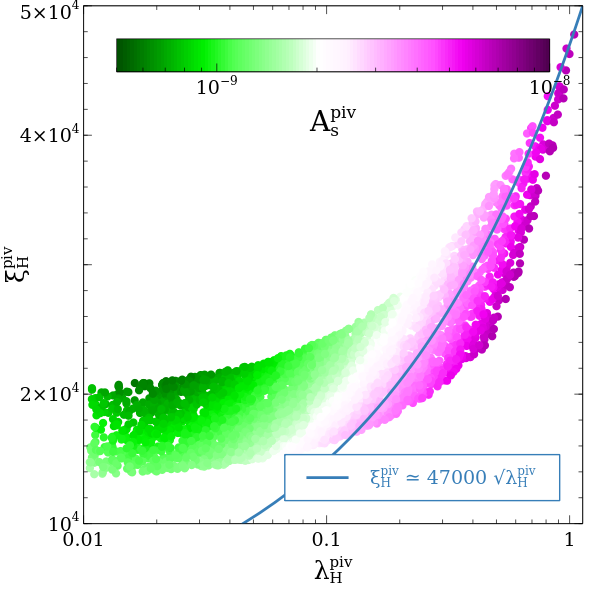}
\includegraphics[angle=0,height=7.5cm,width=7.5cm]{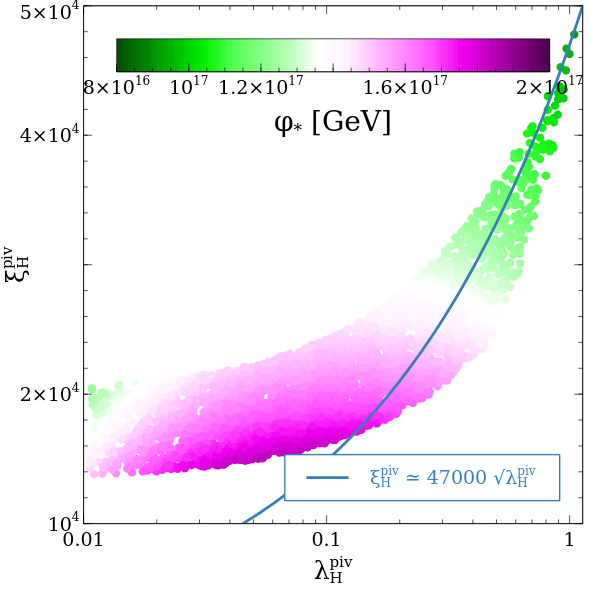}
\caption{Upper LP shows the scatter plot in the $M_{h_2}-\sin\theta$ plane, where the color bar shows the different values of the scalar-to-tensor ratio, whereas variation in the $g_{D} - \Lambda^{\text{inf}}_{H}$ plane has been shown in the upper RP. The colour bar in the RP shows the different values of the curvature power spectrum. The parameter is the inflation condition at the pivot scale, defined in the text. In the lower LP and RP,
we have shown the scatter plots in the $\lambda^{\text{piv}}_{H}-\xi^{\text{piv}}_{H}$ plane and 
the color variation shows the different values of $A^{\text{piv}}_{s}$ and pivot 
scale $\phi_{*}$ in the LP and RP, respectively.} 
\label{scatter-plot-2}
\end{figure}

In Fig. \ref{scatter-plot-2}, we have shown the scatter plots in the 
$M_{h_2}-\sin\theta$, $g_{D}-\Lambda^{\text{\text{inf}}}_{H}$ and 
$\lambda^{\text{piv}}_{H}-\xi^{\text{piv}}_{H}$ planes after satisfying all the constraints
relevant to the Higgs inflation and the DM observables. In the upper LP, we have shown the variation in 
the $M_{h_2}$-$\sin\theta$ plane and the colour variation
shows the values of the tensor to scalar ratio $r_{\text{piv}}$ at 
the pivot scale. We can see
the $r_{\text{piv}}$ value is always less than the current limit 
obtained from the Planck data as depicted in Eq. \ref{As-ns-r-label}.
In general, $r_{\text{piv}}$ is proportional to $\epsilon$, which is proportional to
quartic coupling $\lambda_{H}$, and the colour bar is depicting this behaviour.
In the RP, we have shown the variation in the $g_{D} - \Lambda^{\text{\text{inf}}}_{H}$, 
where $\Lambda^{\text{\text{inf}}}_{H} = \lambda_{HD}-2 \lambda_{H} 
\frac{\xi_{D}}{\xi_{H}}$, plane and the colour vaiation shows the different values
of $A^{\text{piv}}_{s}$. We see linear correlation between $\Lambda^{\text{\text{inf}}}_{H}$ and $g_{D}$
which happen mainly because $\lambda_{HD} \propto \frac{1}{v^2_{D}} \propto 
g^2_{D}$. We can see from the plot that after demanding the 
inflation conditions we always have 
$\Lambda^{\text{\text{inf}}}_{H} > 0$ which is needed for the inflation to happen along the Higgs 
field direction. Moreover, we can see $A_s^{\text{piv}}$ is distributed all over 
the space so we can have the correct value of the $A_s^{\text{piv}}$ obtained from the 
Planck data for any value of the gauge coupling. In the lower two plots,
we have shown the scatter plots in the $\lambda_{H}^{\text{piv}} - \xi^{\text{piv}}_{H}$
plane where the colour variation shows the different values of $A_s^{\text{piv}}$
and pivot scale $\phi_{*}$ in the LP and RP, respectively. We can see a 
nice correlation in the plane and the blue line displays the $\lambda_{H}$
and $\xi_H$ values at the pivot scale which give the correct value of
$A^{\text{piv}}_s$ obtained from the Planck data.

\begin{figure}[h!]
\centering
\includegraphics[angle=0,height=7.5cm,width=7.5cm]{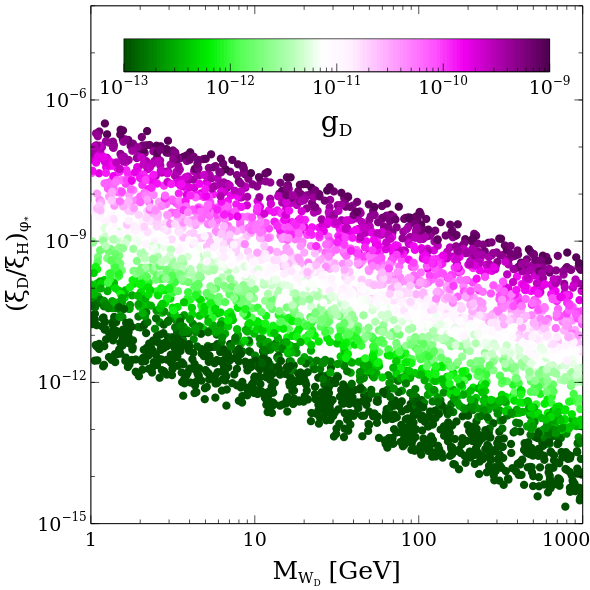}
\includegraphics[angle=0,height=7.5cm,width=7.5cm]{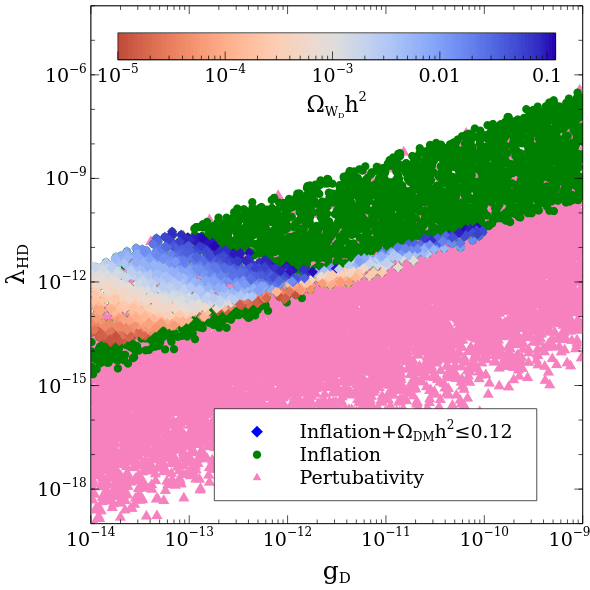}
\caption{LP shows the scatter plots in the $M_{W_{D}} - (\xi_{D}/\xi_{H}){\phi{*}}$ plane, where the colour bar shows the different values of the gauge coupling
$g_D$. RP shows the allowed region in the $g_{D}-\lambda_{HD}$ plane after implementing the bounds from perturbativity, inflation, and the upper 
allowed range on the DM relic density.} 
\label{scatter-plot-3}
\end{figure}

In the LP of Fig. \ref{scatter-plot-3}, we have shown scatter plots
in the $M_{W_D} - (\xi_{D}/\xi_{H})_{\phi_{*}}$ plane 
where $\phi_{*}$ is the pivot scale and the colour variation shows 
the different values of the gauge coupling $g_D$. In the
RP, we have shown in the $g_{D}-\lambda_{HD}$ plane after implementing the 
successive bounds coming from perturbativity, inflation and the upper bound on 
DM relic density. 
In the LP, we can see $(\xi_{D}/\xi_{H})_{\phi_{*}}$
varies inversely with the gauge boson mass $M_{W_D}$ and 
linearly with the gauge coupling $g_D$. This is because during the 
running of $\xi_D$ which is linearly proportional to quartic coupling
and depends on the gauge coupling and gauge boson mass as, 
$\lambda_{HD} \propto \frac{g_{D}}{M_{W_D}}$. We should note that we started with
$\xi_{D} = 0$ at the top-quark pole mass but it has been generated at the pivot 
scale due to the effect of the loop mediated diagrams in the beta function. 
In the RP,
we have shown the variation in the $g_{D}-\lambda_{HD}$ plane after applying the 
consecutive constraints coming from perturbativity (brick coloured triangle points),
inflation (green circle points) and
upper bound on DM density (diamond-shaped points with colour variation bar). 
The brick colour region represents the allowed region
after applying the perturbativity bound as listed in 
Eq. (\ref{pertubativity-bound}), green 
points are obtained when we apply the inflation conditions as listed 
in section \ref{inflation}, along with the bounds on the inflation observables from Planck data and finally, the colour varied points are 
obtained after applying the allowed range of the DM density 
as described in Eq. \ref{DM-density-range}. 
After applying the DM density bound, 
the allowed region significantly reduced because the other points overproduced 
the DM due to the larger values of the gauge coupling. We see a sharp allowed region 
in the points for $g_{D} > 2\times 10^{-12}$ which mostly comes from the 
higher values of the DM mass where decay is not possible and mainly from the 
gauge bosons annihilation including gluon and photon{}
\footnote{It is to be noted that we have considered the starting temperature
for DM production 
at $T_{\text{ini}} = 1.5$ TeV in \texttt{micrOMEGAs}. 
The infrared-dominated freeze-in production
is independent of $T_{\text{ini}}$, but the ultraviolet freeze-in contribution,
which mainly arises from gauge boson annihilations, depends on $T_{\text{ini}}$.
In our scenario, electroweak symmetry breaks at a low temperature,
after which mixing among the Higgses is generated, enabling DM production
from gauge boson annihilation. Therefore, to accommodate higher values of $T_{\text{ini}}$,
electroweak symmetry breaking must occur at a high scale,
as studied in Ref. \cite{Baldes:2018nel}, 
so that the required Higgs mixing can be realised.}. The colour variation in the 
points also depict that if we consider the 100\% DM coming from the present 
DM (blue region), then a very narrow region is allowed combinedly after 
incorporating both inflation and DM bounds. 
The larger region can be considered allowed when we have multi-component DM
and the rest of the contribution comes from the other DM components.

\begin{figure}[h!]
\centering
\includegraphics[angle=0,height=7.5cm,width=7.5cm]{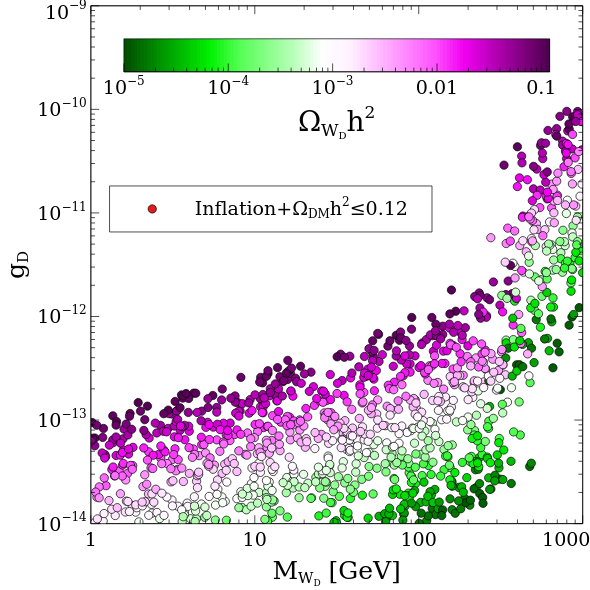}
\includegraphics[angle=0,height=7.5cm,width=7.5cm]{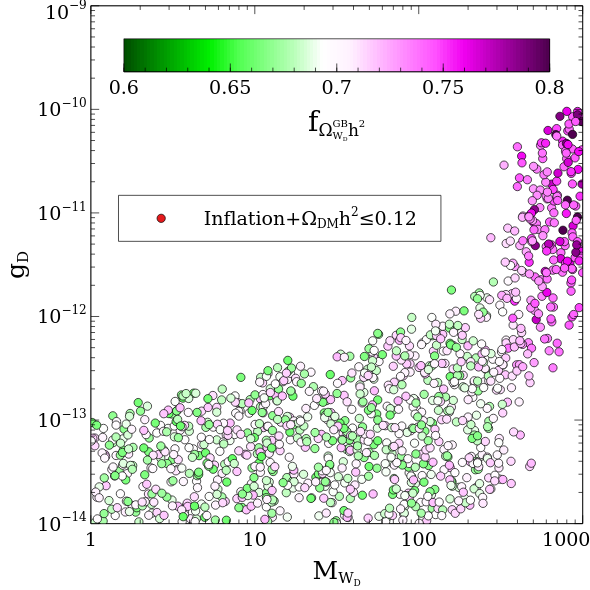}
\caption{LP and RP show the allowed region in the $M_{W_D}-g_{D}$ plane 
after incorporating all the relevant bounds. The colour bar in the LP shows 
the variation of DM density, whereas the colour bar in the RP shows the 
contribution of gauge boson annihilation to the DM density compared to the other contributions to the DM density.} 
\label{scatter-plot-4}
\end{figure}
In the LP and RP of Fig. \ref{scatter-plot-4}, we have shown the
allowed region after applying the constraints
from inflation and DM density in the $M_{W_D}-g_{D}$ plane. The colour bar
in the LP and RP show the values of the DM relic density and the contribution coming from 
the SM gauge bosons annihilation only, respectively. The production of 
DM does not happen at an early time because there is no mixing between the 
Higgses, hence $W_D$ can not be produced from the annihilation of the SM gauge bosons. 
Therefore, we have considered a lower starting temperature $T_{\text{ini}} = 1.5$ 
TeV to produce 
the DM from the gauge bosons annihilation which is 
$T_{\text{ini}}$ dependent\footnote{Higher values of 
$T_{\text{ini}}$ can be considered following Ref. \cite{Baldes:2018nel}, allowing for dominant dark matter production from gluon and photon annihilations, which is left for future work. }. In the LP, we can see
the variation of DM relic density linearly with the gauge coupling from the
colour variation. We can also see a sharp rise in the gauge coupling for
$M_{W_D} > 500$ GeV where the decay process is kinematically not possible
and annihilations take charge of the DM production which demands larger gauge
coupling. Moreover, for a fixed value of $g_D$, if we go to higher values 
of $M_{W_D}$, then we see a decrement in the DM relic density because DM
production from the Higgses decay is mainly proportional to
 $1/v^2_{D} = g^2_{D}/M^2_{W_D}$. On the other hand, in the RP, we have shown 
 the contribution to the DM density from the gauge boson annihilation. We can
 see $M_{W_D} < 500$ GeV, DM is mainly produced from the decay and gauge 
 bosons contribution is subdominant whereas for the opposite limit, when we do not 
 have decay mode open then it is mostly produced from the gauge bosons annihilation
 including gluon and photon, which appear through one loop processes.

\section{Collider Prospects}
\label{Collider-Prospects}
As shown in Eq. \ref{eqn:scalar_pot_full}, 
after the symmetry breaking, the self-interaction 
of the SM Higgses can be expressed as,
\begin{eqnarray}
\mathcal{L} \supset - \frac{1}{3!} \lambda^{SM}_{3} \kappa_{3} h^3_{1}
- \frac{1}{4!} \lambda^{SM}_{4} \kappa_{4} h^4_{1}\,.  
\end{eqnarray}
where $\kappa_{3,4} = \frac{\lambda^{BSM}_{3,4}}{\lambda^{SM}_{3,4}}$.
The vertices for the three Higgses and four Higgses in the SM and BSM sectors 
can be expressed as,
\begin{eqnarray}
 \lambda^{SM}_{3} &=& -\frac{3 M^2_{h_{1}}}{v}\,,\,\,\, 
\lambda^{SM}_{4} = -\frac{3 M^2_{h_{1}}}{v^{2}}\,, \nonumber \\
 \lambda^{BSM}_{3}& = &-3 \biggl[ 2 v \lambda_{H} \cos^{3}\theta 
- 2 v_{D} \lambda_{D} \sin^{3}\theta - \lambda_{HD} \sin\theta \cos\theta
\left( -v \sin\theta + v_{D} \cos\theta \right) \biggr]\,,\nonumber \\
\lambda^{BSM}_{4} &=& -6 \biggl[ \lambda_{H} \cos^{4}\theta
+ \lambda_{HD} \cos^{2}\theta \sin^{2}\theta 
+ \lambda_{D} \sin^{4}\theta \biggr]
\label{lambda3-lambda4}
\end{eqnarray}
In the collider, there have been intense searches recently for the measurement of the Higgs self-coupling. In this context, the production of double Higgs mainly from the gluon fusion (ggF) and vector boson fusion (VBF) processes has been the focus. The current $\sqrt{s} = 13$ TeV with the data collection of integrated luminosity $\mathcal{L} = 126-139\,\,fb^{-1}$ has put a bound on the Higgs trilinear coupling, which is $-0.4 \leq \kappa_{3} \leq 6.3$ at 95$\%$ C.L., with the central value $\kappa_{3} = 3.0$. 
Moreover, the Higgs quartic coupling is unconstrained in the current LHC setup; therefore, we do not have any bound on $\kappa_{4}$ from the collider.
In the collider, the final states consist of $b \bar b b \bar b$, 
$b \bar b \tau^{+} \tau^{-}$, and $b \bar b \gamma \gamma$ have been used
to constrain the $\kappa_{3,4}$ parameters. There is also a future 
projection on the measurement of the Higgs quartic coupling, mainly from the measurement of the Higgs trilinear coupling at the HL-LHC with the data collection of $3\,\,ab^{-1}$ inverse luminosity. The bound on the future Higgs quartic coupling can shrink the allowed region of $\kappa_3$, which is $0.1 \leq \kappa_{3} \leq 2.3$ at 95$\%$ C.L. \cite{Cepeda:2019klc}.
The quartic Higgs coupling coefficient $\kappa_{4}$ can be constrained at the proposed ILC collider \cite{Asner:2013psa}. In Ref. \cite{Liu:2018peg}, they have explored Higgs production and constrained the $\kappa_{3}$ and $\kappa_{4}$ parameters. They put bounds on the parameters as $-27 \leq \kappa_{4} \leq 29$ and $0.89 \leq \kappa_{3} \leq 1.1$, which will exactly predict the SM value. The future ILC has also predicted a large allowed value for $\kappa_{4}$.
There could also be a bound on the Higgs quartic coupling from Higgs self-interaction $h_{1} h_{1} \rightarrow h_{1} h_{1}$, which translates as
\begin{eqnarray}
|\kappa_{4}| \leq \frac{16 \pi v^{2}}{3 M^2_{h_{1}}} = 65,,
\end{eqnarray}
which is very relaxed compared to the coefficients of the Higgs trilinear coupling.
Finally, the proper measurement of the Higgs self-coupling will be very insightful in understanding electroweak symmetry breaking more concretely. In the 
present work, we have also mentioned about the prediction of 
Higgs inflation. 

\begin{figure}[t!]
    \centering
    \includegraphics[angle=0,height=8.5cm,width=10.5cm]{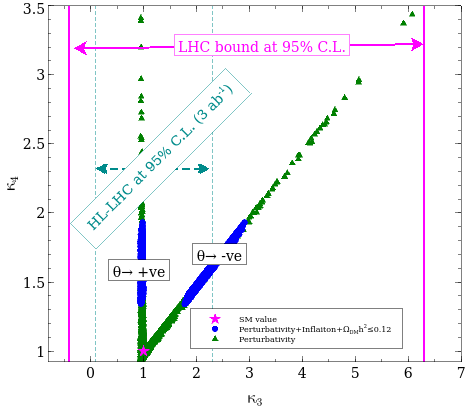}
    \caption{Allowed region in the $\kappa_{3}-\kappa_{4}$ plane after applying
    the bounds from the perturbativity, inflation and upper bound on the DM density.
    In the future, HL-LHC has the possibility to explore the region which is relevant
    for inflation after taking into account the two-loop $\beta-$function
    in the running.}
    \label{quartic-coupling}
\end{figure}
In Fig. \ref{quartic-coupling}, we have shown the allowed parameter regime in the $\kappa_{3}-\kappa_{4}$ plane after applying the successive bounds from perturbativity, inflation, and the maximum allowed DM density. In the figure, we have two types of variation in $\kappa_{3}-\kappa_{4}$ plane depending on the choice of the mixing angle $\theta$. The vertical points around $\kappa_{3} = 1$ are for the positive mixing angle shown in Eq. \ref{lambda3-lambda4}, whereas the diagonal points represent when $\theta$ is replaced by $-\theta$ in Eq. \ref{lambda3-lambda4}.
As can be seen, when we apply the perturbativity bounds, represented by the green triangle points, we see a large allowed region in the plane. We can see for $\theta$ positive value $\kappa_3$ is always around the SM value for the increment of $\kappa_4$ but when $\theta$ is negative we see a nice 
correlation between $\kappa_3$ and $\kappa_4$ because all the relevant parameters are positive in the $\kappa_{3}$ and $\kappa_{4}$ (replacing $\theta\rightarrow -\theta$ in Eq. \ref{lambda3-lambda4}).
Once we apply the inflation bounds, we see that only the blue circle points are allowed, which are away from the star points representing the predicted value by the SM. This difference is mainly due to the need for the quartic Higgs coupling $\lambda_{H} \gtrsim 0.18$; otherwise, the Higgs quartic coupling becomes negative at the high scale, and the potential becomes metastable.
We have also simultaneously used the upper bound on the allowed DM density, and it has no effect on shrinking the allowed region in the $\kappa_{3}-\kappa_{4}$ plane.
In the plots, we see magenta lines, which are the present bounds from the ATLAS detector at LHC for $\sqrt{s} = 13$ TeV and correspond to an integrated luminosity of $126-139$ $fb^{-1}$ data at 95 $\%$ C.L. \cite{ATLAS:2022kbf, ATLAS:2021fet, ATLAS:2022jtk}. The cyan dashed lines represent the allowed range from the HL-LHC at 95$\%$ C.L. with the accumulation of $3\,ab^{-1}$ data. We can see that HL-LHC can explore part of the region that is essential for inflation when we consider $\theta$ value negative, but no as such restrictions for $\theta$ positive value because $\kappa_4$ is difficult to probe at the collider and the bound is weak.
The SM value is represented by the magenta star point, which lies away from the values allowed by inflation in the 
$\kappa_4$ direction for positive $\theta$. However, for negative $\theta$, both 
$\kappa_3$  and $\kappa_4$ deviate from the SM value.
So, with the measurement of the Higgs quartic coupling at HL-LHC in the future, we can validate the fate of the Higgs inflation scenario. If we confirm the SM-predicted value for $\kappa_{3,4}$, then we can directly rule out the Higgs inflation scenario.

\section{Conclusion}
\label{conclusion}
In the present work, we have studied inflation and DM together in the same setup, considering the effect of one sector on the other sector. 
These two phenomena occur at 
different energy scales, so we have used renormalisation group running of the couplings to connect them.
In this context, we have extended the SM symmetry by introducing an abelian 
dark gauge symmetry with its gauge boson serving as the DM candidate
and we have expanded the particle spectrum by adding a dark singlet scalar. 
Once the singlet scalar acquires a spontaneous VEV, then the DM gets mass. 
DM is stabilised by introducing charge conjugation symmetry in the dark sector.
For inflation, we have considered the SM Higgs as the inflaton, which has 
a non-minimal coupling with the Ricci scalar 
\footnote{We plan to further explore dark Higgs inflation in the same setup in 
future work.}. 
When transforming from the Jordan frame to the Einstein frame via a conformal transformation of the metric, we obtain a Starobinsky-type potential in 
the Einstein frame. 
This potential is required for the slow-roll evolution of the inflaton field 
during the inflation in early Universe.
We have computed all inflationary observables at the pivot scale 
or Horizon exit scale,
incorporating the running effects of couplings. These observables include 
the spectral index, tensor-to-scalar ratio, amplitude of curvature power spectrum
and the number of e-folds. 
We found parameter ranges that agree with the CMB constraints obtained from Planck data.
After implementing both inflation and DM constraints, we discovered 
interesting correlations among the model parameters, particularly a very narrow allowed region in the $\sin\theta - M_{h_2}$ plane which fixes the Higgs quartic coupling 
in the range $0.18-0.25$. 
We also found strong correlations in the $g_{D} - \lambda_{HD}$ plane.
To satisfy Higgs inflation conditions, we set the dark Higgs non-minimal coupling 
$\xi_D = 0$ at the top-quark pole mass, even though it becomes nonzero at 
higher scales. This choice (or selecting a very small value) was necessary because all associated quartic couplings became extremely feeble due to the FIMP DM consideration. Otherwise, inflationary conditions could not be satisfied.
For DM production, we considered the freeze-in mechanism which requires gauge couplings 
in the feeble regime. This also results in feeble quartic couplings except for $\lambda_H$. In the DM production by annihilation processes, we incorporated gluon and photon couplings with Higgses generated at the loop level in \texttt{FeynRules} to get the model file 
for \texttt{micrOMEGAs}. This loop contribution of gluon and photon provides a 
moderate effect on DM production
as it depends on the initial temperature ($T_{\text{ini}}$) of DM production. 
However, $T_{\text{ini}}$ cannot be taken too high as that would result in a 
vanishing Higgses mixing angle.
We found that in regions where decay is kinematically forbidden, annihilation processes can fully account for DM production, primarily dominated by SM gauge bosons annihilation. 
When imposing the upper bound on DM relic density, the allowed region in the $g_D - \lambda_{HD}$ plane shrinks significantly becoming even narrower if we assume 100\% 
of the observed DM density comes from $W_D$.
Additionally, we explored the collider aspects of the present study by analysing the parameters $\kappa_3$ and $\kappa_4$ which are associated with the Higgs trilinear and quartic vertices. After incorporating both inflation and DM constraints, we found that the allowed range of $\kappa_{3,4}$ deviates significantly from the SM values $(1,1)$.
Finally, this study represents an effort to connect inflation and DM by evolving relevant parameters at different energy scales using renormalisation group running. We have found a nontrivial effect of DM physics on inflation.
\section{Acknowledgements}
SK acknowledges the visit to IACS, Kolkata, during which this work was initiated.
SK thanks Hyun Min Lee and Jinsu Kim for the fruitful discussion. 
The research is supported by Brain Pool program funded by the Ministry of Science and
ICT through the National Research Foundation of Korea (RS-2024-00407977) and Basic Science
Research Program through the National Research Foundation of Korea (NRF) funded by the
Ministry of Education, Science and Technology (NRF-2022R1A2C2003567). For the numerical
analysis, we have used the Scientific Compute Cluster at GWDG, the joint data center of Max
Planck Society for the Advancement of Science (MPG) and University of G\"{o}ttingen. AT thanks IACS for financial support during the early stages of the project. AT also thanks HIP, Finland, for the funded visit. AT also acknowledges IFJ-PAN, Poland, for financial support.

\newpage
\appendix
\section{Appendix}

\subsection{Renormalisation group equations}
\label{apdx:RGEs}
The beta functions after considering two-loop diagrams are given by
\begin{align}
    (4\pi)^2\beta_{g_1} &= \frac{81+s_H}{12}g_1^3 \,,\\
    (4\pi)^2\beta_{g_2} &= -\frac{39-s_H}{12}g_2^3 \,,\\
    (4\pi)^2\beta_{g_3} &= -7g_3^3 \,,\\
    (4\pi)^2\beta_{g_D} &= \frac{s_D}{3}g_D^3 \,,\\
    (4\pi)^2\beta_{y_t} &= y_t\left[
    \left(\frac{23}{6}+\frac{2}{3}s_H\right)y_t^2
    -\left(
    8g_3^2 + \frac{17}{12} g_1^2 + \frac{9}{4}g_2^2
    \right)
    \right]
    \,,\\
    (4\pi)^2\beta_{\lambda_H} &=
    6(1+3s_H^2)\lambda_H^2 + \frac{1+s_D^2}{2}\lambda_{HD}^2 - 3g_1^2\lambda_H
    -9g_2^2\lambda_H 
    \nonumber\\
    &\quad
    + \frac{3}{8}g_1^4 + \frac{3}{4}g_1^2g_2^2
    +\frac{9}{8}g_2^4 + 12\lambda_H y_t^2 - 6y_t^4
    \,,\\
    (4\pi)^2\beta_{\lambda_D} &=
    2(1+9s_D^2)\lambda_D^2 + \frac{3+s_H^2}{2}\lambda_{HD}^2
    - 12g_D^2\lambda_D + 6g_D^4
    \,,\\
    (4\pi)^2\beta_{\lambda_{HD}} &=
    6(1+s_H^2)\lambda_H\lambda_{HD} + 2(1+3s_D^2)\lambda_D\lambda_{HD}
    +4s_Hs_D\lambda_{HD}^2
    \nonumber\\
    &\quad
    - \frac{3}{2}g_1^2\lambda_{HD}
    - \frac{9}{2}g_2^2\lambda_{HD} 
    - 6g_D^2\lambda_{HD}
    + 6\lambda_{HD}y_t^2
    \,,
\end{align}
where suppression factors $s_H$ and $s_D$ are given by
\begin{align}
    s_H = \frac{1+\xi_H h^2/M_{\rm P}^2}{1+(1+6\xi_H)\xi_H  h^2/M_{\rm P}^2}
    \,,\quad
    s_D = \frac{1+\xi_D \phi^2/M_{\rm P}^2}{1+(1+6\xi_D)\xi_D  \phi^2/M_{\rm P}^2}
    \,.
\end{align}
Without the nonminimal coupling, the suppression factor equals unity.
The beta functions for the nonminimal couplings are given by:
\begin{align}
    (4\pi)^2\beta_{\xi_H} &=
    \left[
    6(1+s_H)\lambda_H - \frac{3}{2}(g_1^2+3g_2^2) + 6y_t^2
    \right]\left(
    \xi_H + \frac{1}{6}
    \right)
    +(1+s_D)\lambda_{HD}\left(
    \xi_D + \frac{1}{6}
    \right)
    \,,\\
    (4\pi)^2\beta_{\xi_D} &=
    \left[
    2(1+3s_D)\lambda_D - 6g_D^2
    \right]\left(
    \xi_D + \frac{1}{6}
    \right)
    + (3+s_H)\lambda_{HD}\left(
    \xi_H + \frac{1}{6}
    \right)
    \,.
\end{align}
Finally, the anomalous dimensions take the form,
\begin{align}
    (4\pi)^2\gamma_H &=
    -\frac{3}{4}g_1^2 -\frac{9}{4}g_2^2 + 3y_t^2
    \,,\\
    (4\pi)^2\gamma_D &=
    -3g_D^2
    \,.
\end{align}

\end{document}